\documentclass{aa}         
\usepackage{graphicx}
\usepackage{txfonts}
\usepackage{natbib}

\newcommand\oao{OAO\,1657$-$415}

\begin{document}
   \title{\emph{INTEGRAL} observations of the variability of \oao\thanks{Based
        on observations with \emph{INTEGRAL}, an ESA project with instruments
        and science data centre funded by ESA member states (especially the PI
        countries: Denmark, France, Germany, Italy, Switzerland, and Spain), the
        Czech Republic, and Poland and with the participation of Russia and the US.}}

   \author{J. Barnstedt
          \inst{1}
       \and
          R. Staubert
          \inst{1}
       \and
          A. Santangelo
          \inst{1}
       \and
          C. Ferrigno
          \inst{2}
       \and
          D. Horns
          \inst{1}
       \and
          D. Klochkov
          \inst{1}
       \and
          P. Kretschmar
          \inst{3}
       \and
          I. Kreykenbohm
          \inst{1,5}
       \and
          A. Segreto
          \inst{2}
       \and
          J. Wilms
          \inst{4}
          }

   \offprints{J. Barnstedt}

   \institute{Kepler Center for Astro and Particle Physics,
              Institut f\"ur Astronomie und Astrophysik, Universit\"at T\"ubingen,
              Sand 1, 72076 T\"ubingen, Germany\\
              \email{barnstedt@astro.uni-tuebingen.de}
         \and
              INAF IFC-Pa, via U. La Malfa 153, 90146 Palermo, Italy
         \and
              ESA, ESAC, P.O. Box 78, 28691 Villanueva de la Ca\~nada, Spain
         \and
              Dr. Karl Remeis-Sternwarte, Astronomisches Institut, Universit\"at Erlangen-N\"urnberg, Sternwartstr. 7, 96049 Bamberg, Germany
         \and
              INTEGRAL Science Data Centre, Chemin d'\'Ecogia 16, 1290 Versoix, Switzerland
             }

   \date{Received ; accepted }

  \abstract
   {The Galactic Plane Scan (GPS) was one of the core
   observation programmes of the \emph{INTEGRAL} satellite. The highly
   variable accreting pulsar \object{\oao } was frequently observed
   within the GPS.}
   {We investigate the spectral and timing properties of \oao\ and
   their variability on short and long time scales in the energy
   range $6-160$~keV.}
   {Using standard extraction tools and custom software for
   extracting \emph{INTEGRAL} data we analysed energy-resolved light
   curves with a time resolution of one second -- mainly data of the
   ISGRI instrument. We also analysed phase-averaged broad band
   spectra -- including JEM-X spectra -- and pulse-phase resolved
   spectra of ISGRI.}
   {During the time covered by the \emph{INTEGRAL} observations, the
   pulse period evolution shows an initial spin-down, which is
   followed by an equally strong spin-up. In combining our results
   with historical pulse period measurements (correcting them for
   orbital variation) and with stretches of continuous observations
   by BATSE, we find that the long-term period evolution is
   characterised by a long-term spin-up overlayed by sets of
   relative spin-down/spin-up episodes, which appear to repeat
   quasi-periodically on a 4.8\,yr time scale. We measure an updated
   local ephemeris and confirm the previously determined orbital
   period with an improved accuracy. The spectra clearly change with
   pulse phase. The spectrum measured during the main peak of the
   pulse profile is particularly hard. We do not find any evidence
   of a cyclotron line, wether in the phase-averaged
   spectrum or in phase-resolved spectra.}
   {}


   \maketitle
%

\section{Introduction}

  \begin{figure*}
\sidecaption
\includegraphics[width=12cm]{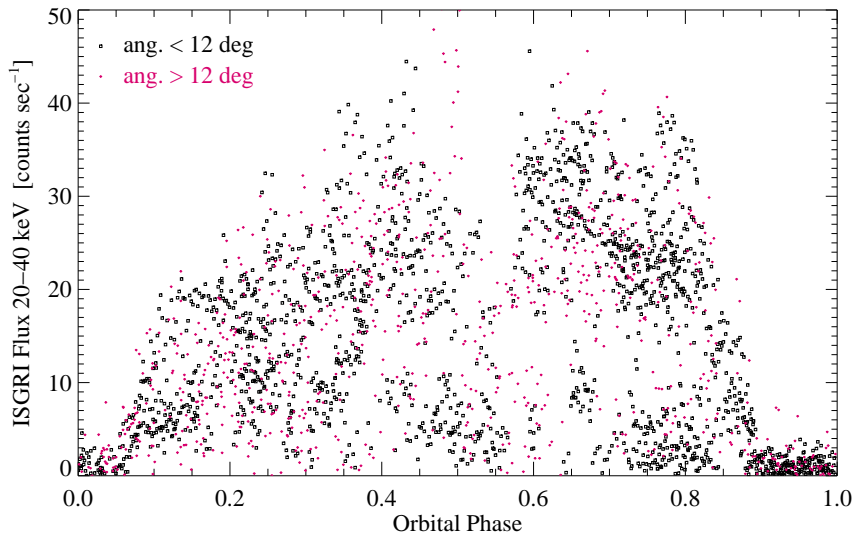}
  \caption{Orbital profile from 2687 individual science
  windows in the ISGRI energy range $20-40$~keV for source
  offset angles $<12\degr$ (black squares) and $12\degr-15\degr$
  (red dots). Though the performance of the IBIS instrument
  significantly decreases for angles larger than $10\degr$, the
  count rate values for offset angles in the range
  $12\degr-15\degr$ are consistent with the count rates for angles
  below $12\degr$. The sharp decrease near phase 0.55 is clearly
  visible.}
\label{FigScatterLightcurve}
\end{figure*}

  The accreting pulsar \oao\ was discovered by
  \citet{1978Natur.275..296P} with the Copernicus satellite.
  \citet{1979ApJ...233L.121W} detected a $38.22$~s pulsation,
  whereas the binary orbit and the eclipsing nature of the high mass
  binary system were discovered by \citet{Chak1993}. The orbital
  period was later improved by \citet{Bild1997} from BATSE
  measurements to $P_\mathrm{orb}=(10\fd44809\pm 0\fd00030)$. As
  \oao\ lies in a heavily absorbed region, an optical counterpart
  could not be identified up to a limit of $V>23$, but an infrared
  counterpart was found by \citet{2002ApJ...573..789C}. The infrared
  properties of \oao\ were found to be consistent with a highly
  reddened B supergiant. Analysing \emph{ASCA} observations
  \citet{Audley2006} found a dust-scattered X-ray halo whose
  evidence decays through the eclipse. Using this halo they
  estimated a distance of $7.1\pm 1.3$~kpc. This is consistent with
  a source distance of $6.4\pm 1.5$~kpc derived by
  \citet{2002ApJ...573..789C} from photometric infrared
  measurements. The pulsar was found to show a long-term spin-up
  from $38.218$~s \citep{1979ApJ...233L.121W} to $37.329$~s
  \citep{2002ApJ...573..789C} with intermediate spin-down periods
  \citep{Bild1997} (see Table~\ref{Tab4HistPeriods}). This
  corresponds to an averaged long-time period derivative of
  $\dot{P}=-1.3\times 10^{-9} s\,s^{-1}$. For a time span of about
  3300 days there are nearly continuous BATSE observations available
  with daily pulse period and flux values. Analysing the spin-up and
  spin-down episodes of BATSE data, \citet{1997A&A...319..515B}
  found that the most natural explanation of X-ray flux and angular
  acceleration fluctuations is the formation of episodic accretion
  disks from stellar wind accretion. \citet{Love1999} developed a
  model for magnetic, propeller driven outflows that cause a rapidly
  rotating magnetised star accreting from a disk to spin-down.
  Another model was introduced by \citet{2006A&A...451..581D} in
  which they assume, in contrast to the model of \citet{Love1999},
  that accretion continues during the (soft) propeller stage. This
  avoids otherwise occurring singularity problems and also leads to
  the fact that no significant changes in the X-ray luminosity occur
  during the spin transitions.

  {The spectral models commonly used for fitting the spectra
  of \oao\ usually include an absorbed power law model for the low
  energy part of the spectrum. For the high energy part an
  exponential cutoff power law model or a power law with high energy
  cutoff is used (see Sect.~\ref{SpectralModeling_Introduction} for
  a definition of these models).

  \citet{Orland1999} searched for cyclotron lines in \emph{BeppoSAX}
  spectra of \oao\ and were able to fit a broad Lorentzian
  line at 30\,keV with a width of 28\,keV, using the exponential
  cutoff model for the continuum. As the width was quite large and
  not well constrained, they fixed the width at 10\,keV, which
  changed the line position to 36\,keV. Using the high energy cutoff
  continuum model, it was found that adding an absorption line did
  not significantly improve the fit. So \citet{Orland1999} concluded
  that the spectrum of \oao\ above 10\,keV can be modelled by a
  series of three power laws with increasing steepness, such that no
  cyclotron line is needed.

  Also, a fluorescence iron line emission was observed in this
  source with several instruments, e.g. \emph{ASCA}
  \citep{Audley2006}, \emph{Chandra}/HETGS
  \citep{2002ApJ...573..789C}, \emph{RXTE} \citep{Baykal2000},
  \emph{BeppoSAX} \citep{Orland1999} and \emph{Ginga}
  \citep{1990PASJ...42..785K}.


\section{Observations and data reduction}

  The satellite \emph{INTEGRAL} \citep{Winkler2003_INTEGRAL} is
  equipped with three instruments for the X- and
  $\gamma$-ray range: the spectrometer SPI (20~keV -- 8~MeV)
  \citep{Vedrenne2003}, the imager IBIS (15~keV -- 10~MeV)
  \citep{Ubertini2003}, and the X-ray monitor JEM-X ($3-60$~keV)
  \citep{Lund2003}. The imager IBIS consists of the low energy CdTe
  detector layer ISGRI ($15-1000$~keV) \citep{Lebrun2003} and the
  high energy CsI detector layer PICsIT ($0.175-10.0$~MeV)
  \citep{Labanti2003}. IBIS has a fully coded field of view (FOV) of
  $9^\circ$ and a partially coded FOV of $19^\circ$ (50\%). In this
  paper we have used ISGRI observations of \oao, which were mainly
  part of the \emph{INTEGRAL} Galactic Plane Scan
  (GPS)\footnote{\emph{INTEGRAL} GPS Team for accreting neutron
  stars (Wilms, Santangelo, Staubert et al., 2004). See also {\tt
  http://pulsar.sternwarte.uni-erlangen.de/gps/ targettable.cgi}. }.
  Being part of the core program of \emph{INTEGRAL}, the GPS
  observations were performed as a sawtooth pattern along the
  accessible part of the Galactic plane with an extension in
  latitude of $\pm10\degr$ \citep{Winkler2003}. A total of 2687
  science windows were available. One \emph{INTEGRAL} science window
  (SCW) is a single observation of about 2000~s, so the total time
  of data acquisition was roughly $\sim 5$~Ms. The data used range
  from revolution 36 (MJD 52\,668) up to revolution 463 (MJD
  53\,946).

  Spectra were obtained also from JEM-X data. Due to a smaller FOV
  (fully coded $4\fdg{}8$, partially (50\%) coded $7\fdg{}5$), only
  41 science windows (SCWs) from JEM-X1 and 46~SCWs from
  JEM-X2 were available (selected from the SCWs used for ISGRI
  data, but with the angular distance of the centre of the FOV to
  the source restricted to $<4\fdg{}5$). These were JEM-X2
  observations of February and March 2003 and JEM-X1 observations
  from August 2004 to April 2005. As these 87~SCWs are just a small
  fraction of the total SCWs available from ISGRI
  observations, only phase-averaged spectra were produced
  (OSA\,6.0), but no timing analysis was done for JEM-X data.

  Generally, the observational data have been reduced using the
  Offline Scientific Analysis (OSA) software v.~6.0 provided by the
  \emph{INTEGRAL} Science Data Centre, ISDC \citep{Cour2003}. For
  generating spectra and pulse profiles from ISGRI data
  (spectral timing), an alternative software provided by
  the IASF Palermo\footnote{ \tt
  http://www.pa.iasf.cnr.it/$\sim$ferrigno/ INTEGRALsoftware.html
  \label{fn_palermo}} \citep{Ferrigno2007} was used.

  For the timing analysis (including spectral timing
  analysis) we selected all observations for which \oao\ was within
  $12\degr$ from the centre of the FOV (this is recommended by ISDC
  since the reliability of the flux determination is reduced when
  the source is located in the outer parts of the partially coded
  FOV of ISGRI). For all existing science windows we determined the
  orbital phase and selected only data in the orbital phase interval
  from 0.1 to 0.9, thus excluding data during the eclipse.

  Furthermore, we included only those science windows for
  which the count rate is above 15\,s$^{-1}$ in the energy interval
  $20-40$\,keV, corresponding to the ``high state'' (see below). The
  reason for this is that the timing analysis (the determination of
  the pulse period and in particular the determination of the
  pulse phase), needs data with a sufficient signal to noise ratio to
  yield significant results. All above mentioned restrictions left
  842 science windows out of 2687 available (see also
  Fig.~\ref{FigScatterLightcurve}, which shows data from all
  available SCWs). Our results are therefore applicable to
  this ``high state'' only.

\begin{figure}
\includegraphics[width=1.0\columnwidth]{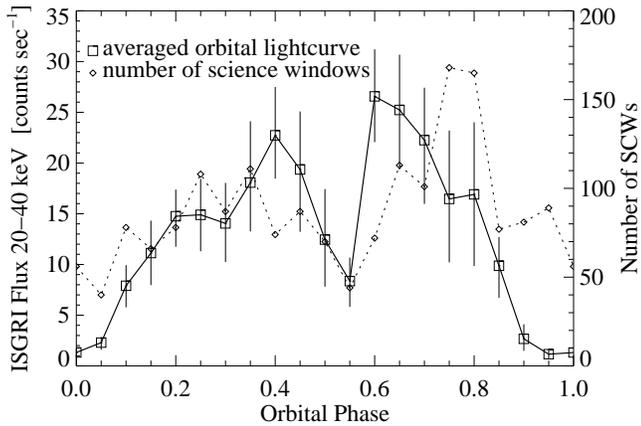}
  \caption{Averaged orbital profile in the ISGRI energy band
  $20-40$\,keV. This profile was calculated from 1797 science
  windows for which the angular distance to the telescope axis was
  less than $12\degr$ (see text for description of error bars). The dashed
  line connects values giving the number of science windows (right
  axis) used for the respective averages.}
\label{FigOrbitalLightcurve}
\end{figure}

\section{Timing analysis}
\label{Timing_analysis}
\subsection{The orbital profile}

  Figure~\ref{FigScatterLightcurve} shows the mean count rates from
  2687 science windows in the ISGRI energy band $20-40$\,keV as a
  function of orbital phase, which was calculated using the
  ephemeris of \citet{Bild1997} ($T_{90}$ = MJD~48\,515.99 and
  $P_\mathrm{orb}$ = 10.448\,09\,d). The large scatter in this
  figure reflects the strong intrinsic variability of \oao.
  Fig.~\ref{FigOrbitalLightcurve} shows the averaged orbital
  profile, which was calculated from 1852 science windows (MJD
  $52\,668-53\,946$) for which the angular distance of \oao\ from the
  centre of the FOV was less than $12\degr$. The dashed line shows
  the number of science windows for each data point. In general, we
  confirm the asymmetric shape of the profile as it was
  measured by \citet{2006ApJS..163..372W} with ASM on board
  \emph{RXTE}. In contrast to these measurements, the
  \emph{INTEGRAL} observations show a remarkable minimum at
  phase~0.55, which is visible also in the ASM data averaged for the
  time range MJD~$52\,668-53\,946$ (Fig.~\ref{FigOrbitalLightcurveASM}).
  The data presented by \citet{2006ApJS..163..372W} cover a time
  range of roughly 8.5~years ($\sim$~MJD~$50\,160-53\,240$), which is
  much wider than the time range we analysed. We conclude therefore
  that the dip seen in our data might be a temporary phenomenon.

  The lack of data points around $10-15$~counts\,s$^{-1}$ (cps) at
  orbital phases above 0.4 apparent in
  Fig.~\ref{FigScatterLightcurve} may suggest the existence of
  a bimodal intensity distribution, defining a
  ``high state'' and a ``low state'' of this source.

  The formal uncertainties of the count rates in
  Fig.~\ref{FigOrbitalLightcurve} are negligible, but an
  unpredictable scatter is introduced by the fact that only in rare
  cases the continuous measurements span a complete orbit. Due to
  the intrinsic variability of the source the available data points
  do not represent a homogeneously averaged profile. The
  number of data points (Science Windows) per phase bin used to
  calculate the average flux varies between 40 and nearly 170. To
  evaluate the influence of the non-homogeneity of this sampling, we
  calculated profiles with a constant number of 30 SCWs per phase
  bin, randomly chosen from the available SCWs. We calculated
  10\,000 of these profiles and plotted the range between minimum
  and maximum values for each phase bin as ``error bars'' to
  Fig.~\ref{FigOrbitalLightcurve}. A discussion is given in
  Sect.~\ref{Summary_and_discussion}.

\begin{figure}
\includegraphics[width=1.0\columnwidth]{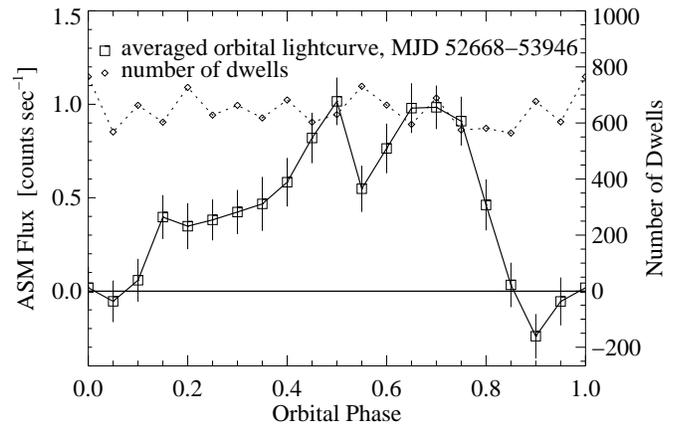}
  \caption{Averaged orbital profile (ASM, $2-10$\,keV) plotted from
  quick-look results provided by the ASM/RXTE team. This curve is
  calculated from 12830 single dwell values (sum band intensity)
  covering the used \emph{INTEGRAL} observation period from MJD
  $52\,668-53\,946$. The error bars are quadrature averages of the
  flux errors. The dashed line shows the number of dwells per data
  point (right axis).}
\label{FigOrbitalLightcurveASM}
\end{figure}

\begin{figure}
\includegraphics[width=1.0\columnwidth]{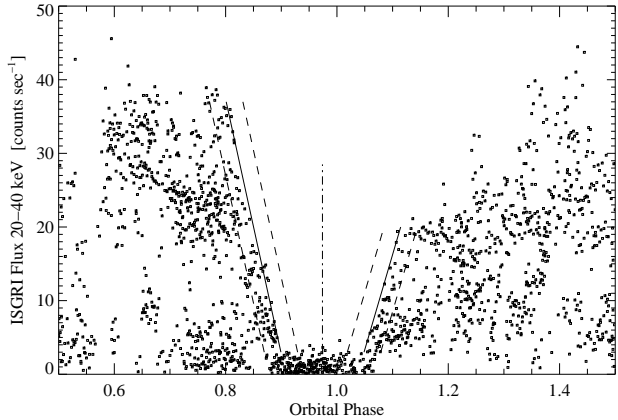}
  \caption{Determination of the centre of eclipse by fitting
  straight lines to ingress and egress. The dashed lines denote the
  width of the strips (0.03 phase units), which were used to
  determine the number of data points falling into the strips left
  and right to the fitted straight lines (see text). Only data
  points from SCWs with offset angles $<12\degr$ were used for this
  analysis (see Fig.~\ref{FigScatterLightcurve}).}
\label{FigEclipse}
\end{figure}

  Using the scatter plot shown in Fig.~\ref{FigScatterLightcurve} we
  have determined the properties of the eclipse. The data points
  left and right of the apparent eclipse define a rather sharp
  boundary in phase space (orbital phase vs. count rate), leaving
  only a few individual outlying points. We fitted these boundaries
  by straight lines by maximising the differences in data points
  that fall into strips of predefined widths left and right to these
  lines (see Fig.~\ref{FigEclipse}). For ingress we fitted a line
  from 37~cps down to the maximum eclipse intensity of $\sim 3$~cps
  and the egress line from 3~cps to 20~cps. The strip widths were
  varied from 0.01 to 0.05 in phase units. No significant change of
  the fit parameters was observed for strip widths in the range
  $0.02-0.04$. We therefore used the average values obtained from
  the fits with 3 widths of 0.02, 0.03 and 0.04. The intersections
  of the ingress and egress lines with the bottom count rate are at
  orbital phases 0.899(5) and {1.052(5) (corresponding to
  MJD~52\,662.84(5) and MJD~52\,664.44(5), respectively), such that
  we set the centre of eclipse to phase 0.976(5)
  (MJD~52\,663.64(5)), and the half width of the full eclipse to
  0\fd80(3) (or $27\fdg{}6\pm 1\fdg{}0$). The MJD values are
  determined in such a way that they fall into the last orbit before
  our observations. The straight lines for ingress and egress are
  described by slopes (change of count rates) of 34~cps\,d$^{-1}$
  and 28~cps\,d$^{-1}$, respectively, making the ingress faster than
  the egress by a factor of 1.2. We note that the above description
  gives upper limits to the speed of ingress or
  egress, individual events may be slower.


  \begin{table}

      \caption[]{Pulse periods of \oao\ as measured during the
      \emph{INTEGRAL} GPS in 11 observation groups. The
      date is the centre of the observation time interval.}

      \label{INTEGRAL_Periods}
      \centering
      \begin{tabular}{r l l l l}
      \hline \hline
             Grp.     & Date       & Range        & Period       &  $\dot{P}$\\
                      & [MJD]      & [d]          & [s]          &  [$10^{-9}~s\,s^{-1}$] \\
      \hline
             1        &  52672.03  & $\pm$0.67    & 37.1467  (1) &                    \\
             2        &  52711.48  & $\pm$12.87   & 37.1645  (5) &  $2.9\pm1.0$       \\
             3        &  52859.94  & $\pm$0.18    & 37.2028  (5) &                    \\
             4        &  52869.95  & $\pm$0.91    & 37.2005  (5) &                    \\
             5        &  52908.42  & $\pm$0.35    & 37.2008  (5) &                    \\
             6        &  53054.92  & $\pm$0.07    & 37.2262 (50) &                    \\
             7        &  53232.53  & $\pm$0.53    & 37.2712  (2) &                    \\
             8        &  53255.48  & $\pm$0.36    & 37.2819  (3) &                    \\
             9        &  53464.74  & $\pm$1.63    & 37.2060 (15) &                    \\
             10       &  53650.10  & $\pm$0.57    & 37.1137  (7) &                    \\
             11       &  53789.58  & $\pm$5.32    & 37.1209  (6) &                    \\
      \hline
      \end{tabular}
  \end{table}

\subsection{Pulsar period analysis}
\label{Pulsar_period}

  The pulse period analysis was performed in steps,
  successively improving the accuracy of the analysis. This
  was necessary because of the strong variation in the pulse period
  over the time covered by the \emph{INTEGRAL} GPS observations. We
  started out with an ISGRI light curve in the $20-200$~keV energy
  range with a time resolution of 1~s (produced by the IASF Palermo
  software$^{\ref{fn_palermo}}$). The bin times were first converted
  to the solar system bary centre and then corrected for binary
  motion using the ephemeris of \citet{Bild1997} ($T_{90}$ =
  MJD~48\,515.99(5), $P_\mathrm{orb}=10\fd448\,09(30)$, $a\sin
  i=106.0(5)$~lt-sec, $e=0.104(5)$ and $\omega=93(5)\degr$). For
  each SCW a period search with epoch folding was performed. This
  led to first estimates of the pulse periods and already clearly
  showed a strong variation with time. As a next step we produced
  pulse profiles combining four to five SCWs. A phase connection
  analysis of these profiles -- by using the mean profile as
  a template and fitting the individual profiles to this template --
  \citep{Deeter1981,Nagase1989,Muno2002} yielded significantly more
  accurate periods. A further refinement was reached using a
  smoothing technique: pulse profiles were generated (with the so
  far best pulse period) for data sets of at least eight SCWs, where
  the start of each data set was shifted by one SCW as compared to
  the start of the previous one. This produced ``running mean pulse
  profiles'' with good photon statistics, from which the final pulse
  periods were again determined by the phase connection technique
  (the loss of statistical independence between consecutive
  pulse profiles does not pose a problem here). The above described
  procedure was performed for each of the eleven groups of
  observations (the groups contained 12 to 273 SCWs). The final
  periods are listed in Table~\ref{INTEGRAL_Periods} and plotted in
  Fig.~\ref{oao_integral}. Due to the short integration
  times it was generally not possible to measure a period derivative
  within one group of observations -- except for group no.~2, for
  which we find a local $\dot{P}$ of $(2.9\pm 1.0)\times
  10^{-9}$~s\,s$^{-1}$ (see Table~\ref{INTEGRAL_Periods}). Phase
  connection between the groups was not possible, as the large
  gaps in time did not allow the unambiguous determination of the
  number of pulsar periods.

\begin{figure}
\includegraphics[width=1.0\columnwidth]{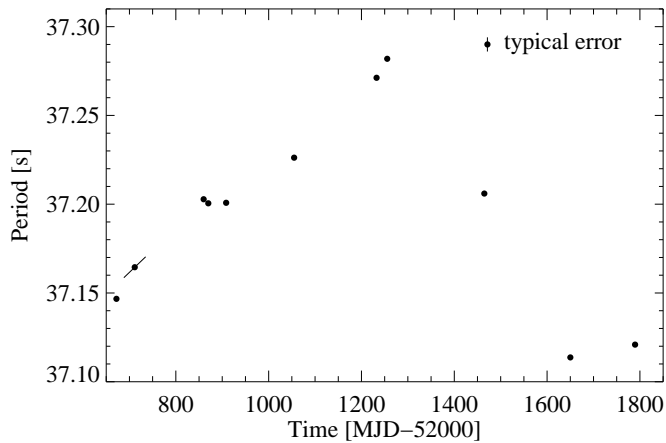}
  \caption{Period evolution of \oao\ as determined in this work
  from \emph{INTEGRAL} ISGRI data (Table~\ref{INTEGRAL_Periods}).
  The typical error of 0.004~s shown here was assumed for periods
  estimated by the epoch folding method. The actual errors for the
  plotted periods estimated by the phase connection method are in
  general about a factor of 10 lower. The $\dot{P}$ value for the
  second data point is indicated by the slope of a short line (whose
  length does not represent the time interval of the evaluated
  observations).}
\label{oao_integral}
\end{figure}

  The period evolution over the time covered by the
  \emph{INTEGRAL} GPS (Fig.~\ref{oao_integral}) is basically a
  triangular function showing an initial spin-down with a mean
  $\dot{P}=+2.5\times 10^{-9}$~s\,s$^{-1}$, followed by a spin-up
  with a mean $\dot{P}=-4.8\times 10^{-9}$~s\,s$^{-1}$. As will be
  shown below, the long-term mean spin-up in \oao\ is
  $\dot{P}=-1.24\times 10^{-9}$~s\,s$^{-1}$, meaning that the values
  measured by \emph{INTEGRAL} during the spin-down and spin-up
  episodes are symmetric with respect to the long-term mean
  value of the period derivative.


  \begin{table*}
      \caption[]{Historic pulsar periods, corrected for binary
                 doppler shift. The errors given for the corrected periods
                 are taken from the errors given for the observed periods
                 (except for the value marked $^{\mathrm{a}}$).}
         \label{Tab4HistPeriods}
         \centering
         \begin{tabular}{l l l c l l}
         \hline \hline
                Date     & Obs. Period & Corr. Period                & Orb. Phase & Instrument              & Reference \\
                {[MJD]}  & [s]        & [s]                          &            &                  & \\
         \hline
                43755.2  & 38.218     & 38.2393 (40)                 &  0.34 & \emph{HEAO A-2}              & \citet{1979ApJ...233L.121W} \\
                44110.2  & 38.019     & 38.0424 (90)                 &  0.32 & \emph{Einstein MPC}          & \citet{1980MNRAS.193P..49P} \\
                45535    & 37.885     & 37.8671 (90) $^{\mathrm{a}}$ &  0.69 & \emph{Tenma GSPC}            & \citet{1984PASJ...36..667N} \\
                47237.5  & 37.747     & 37.7289 (10)                 &  0.63 & \emph{Ginga LAC}             & \citet{1990PASJ...42..785K} \\
                47258.5  & 37.725     & 37.7058 (10)                 &  0.64 & \emph{Ginga LAC}             & \citet{1990PASJ...42..785K} \\
                47589.5  & 37.713     & 37.7355 (30)                 &  0.32 & \emph{Ginga LAC}             & \citet{1990PASJ...42..785K} \\
                47977    & 37.853     & 37.8652 (200)                &  0.41 & \emph{SIGMA}                 & \citet{1991ApJ...366L..23M} \\
                48305    & 37.7897    & 37.7621 (3)                  &  0.81 & \emph{ART-P}                 & \citet{1991IAUC.5342....2S} \\
                48490    & 37.6726    & 37.6707 (15)                 &  0.51 & \emph{SIGMA}                 & \citet{1991IAUC.5342....2S} \\
                50683.954&            & 37.347411 (6) $^{\mathrm{b}}$&       & \emph{RXTE}                  & \citet{Baykal2000}          \\
                50708.5  & 37.39      & 37.3649 (100)                &  0.85 & \emph{ASCA}                  & \citet{Audley2006}          \\
                51950.84 & 37.329     & 37.3019 (200)                &  0.75 & \emph{Chandra HETGS/ACIS-S}  & \citet{2002ApJ...573..789C} \\
         \hline
         \end{tabular}
        \begin{list}{}{}
        \item[$^{\mathrm{a}}$] averaged over 5 days, resulting in an increased error
        \item[$^{\mathrm{b}}$] published value already corrected for binary motion,
                               $\dot{P}=+(4.56\pm 0.13)\times 10^{-9}$~s\,s$^{-1}$
        \end{list}
   \end{table*}

\subsection{Long-term pulse period evolution}
\label{Longterm_period}

  Figure~\ref{oao_longterm} shows the long-term evolution of the
  pulse periods for all data available to us. There are essentially
  three groups of data. First group: the ``historical periods'' as
  taken from the literature and summarised in
  Table~\ref{Tab4HistPeriods}. All except one of these reported
  periods were without binary correction, since most of them were
  determined before the detection of the binary nature of \oao.
  Therefore we have applied the binary correction using the orbital
  parameters of \citet{Bild1997}. Second group: BATSE data as taken
  from public archives\footnote{{\tt
  ftp://legacy.gsfc.nasa.gov/compton/data/batse/
  pulsar/histories/oao1657-415\_8369\_10302.fits.gz
  }}$^,$\footnote{{\tt http://f64.nsstc.nasa.gov/batse/pulsar/data/
  sources/oao1657.html}} and a complementing (and partially
  overlapping) data set provided through private communication by
  Mark Finger. The third group is represented by the data we
  processed from \emph{INTEGRAL} observations presented in this
  paper.

\begin{figure*}
\sidecaption
\includegraphics[width=12cm]{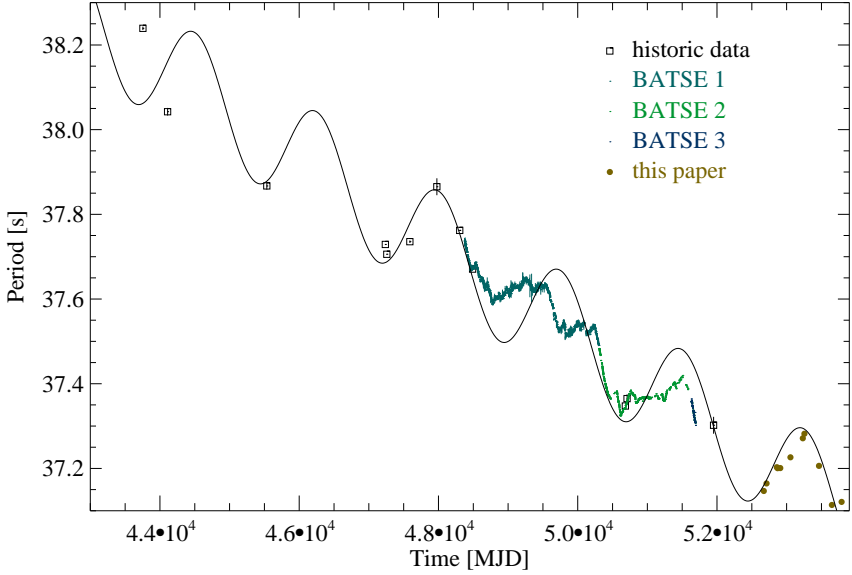}
  \caption{Long-term period evolution of \oao . The historic data
  from the literature (see Table~\ref{Tab4HistPeriods}) were
  corrected for binary motion. The BATSE data are from public
  archive (1,3) and private communication by Mark Finger (2). The
  sinusoid with a period of 4.8~yr, superposed on a linear long-term
  spin-up, is to guide the eye to emphasise a quasi-periodic
  behaviour.}
\label{oao_longterm}
\end{figure*}

  The complete set of data in Fig.~\ref{oao_longterm} is described
  by a mean spin-up rate of $\dot{P}_\mathrm{mean}=-1.24\times
  10^{-9}$~s\,s$^{-1}$. On top of this long-term mean spin-up
  deviations are observed with episodes of strong relative
  ($\dot{P}-\dot{P}_\mathrm{mean}$) spin-up and spin-down. In some
  of the spin-down episodes even the absolute spin-down $\dot{P}$ is
  quite strong (as in the first half of our measurements
  around MJD~53\,000). The appearance in time of the deviations is
  suggestive of a quasi-periodic behaviour. We do not claim a formal
  value for the period, but show in Fig.~\ref{oao_longterm} a model
  modulation by a sine curve with an amplitude of 0.13~s and a
  period of 1750~d (4.8~yr), which reproduces the main features of
  the long-term evolution. One could consider this period
  as the characteristical time for a typical spin-down/spin-up
  cycle.

\subsection{Binary ephemeris}
\label{Binary_ephemeris}

  After having used the binary ephemeris of \citet{Chak1993} and
  \citet{Bild1997} for the pulse period analysis (as discussed in
  Section~\ref{Pulsar_period}), we have analysed the
  non-binary-corrected light curve data from the \emph{INTEGRAL} GPS
  in order to establish our own local ephemeris. We again made use
  of ``running mean pulse profiles'' (from sets of in general eight
  SCWs, the start of each set shifted by one SCW as compared to the
  previous one) and the phase connection techniques. For each of the
  11 observation groups (see Table~\ref{INTEGRAL_Periods}) a local
  ephemeris was established, which -- except for observation 2 --
  are generally not very accurate because the observations are short
  and the intrinsic pulse period shows a strong variation with time
  (see above). Observation~2, however, samples about 2.5 binary
  cycles reasonably densely, such that a fit of the pulse arrival
  time delays (Fig.~\ref{FigPulseDelay}) can be performed and an
  ephemeris established. We find (keeping the orbital elements as
  before): $T_{90}=$~MJD~$52\,663.893(10)$. Adding all other
  observations to the analysis yields the same result (with
  increased uncertainty).

\begin{figure}
\includegraphics[width=1.0\columnwidth]{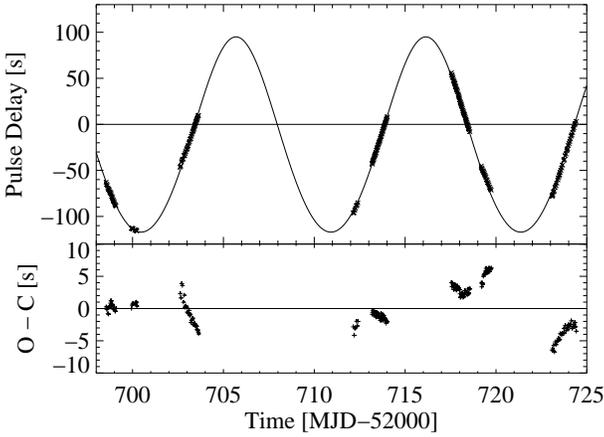}
  \caption{Observed pulse delay times for the second observation
  period, which covers about 2.5 orbital cycles. The solid line
  shows the best fit for the orbital parameters (see
  Sect.~\ref{Binary_ephemeris}). The residuals (observed $-$
  calculated, lower panel) show some variation that may well be due
  to real pulse period variations on time scales shorter than one
  orbital period.}
\label{FigPulseDelay}
\end{figure}

  There is no indication for a secular change of the orbital period.
  Assuming $\dot{P}_\mathrm{orb}=0$, we can combine our new
  ephemeris with values from the past to arrive at a new sidereal
  period. Using, e.g., the $T_{90}$ of \citet{Chak1993} of
  MJD~48\,515.99(5), and dividing the time difference to our
  ephemeris by 397 orbital cycles, we arrive at
  $P_\mathrm{sid}=P_\mathrm{orb}=10\fd448\,12\pm 0\fd000\,13$. Our
  $P_\mathrm{orb}$ is consistent with the value of \citet{Bild1997}
  within our uncertainty, which is reduced by a factor of 2.5 in
  comparison to that of \citet{Bild1997}.
  Table~\ref{OrbitalParameters} shows the current best set of
  orbital parameters.

  We have done the same for the observed times of mid-eclipse: if
  the elapsed time between our mid-eclipse
  (MJD~52\,663.64(5), see above) and that of
  \citet{Chak1993} (their Table~1: MJD~48\,515.897(72)) is divided
  by 397 cycles we find $P_\mathrm{ecl} = 10\fd447\,7\pm
  0\fd000\,3$. The period found from eclipse timing is consistent
  (within uncertainties) with that found using the orbital ephemeris
  determined from pulse timing analysis. We point out that this need
  not be the case: with the finite eccentricity of $e=0.104$ apsidal
  motion is expected, and $P_\mathrm{ecl}$ is only equal to
  $P_\mathrm{orb}$ if $P_\mathrm{ecl}$ is found from averaging over
  one (or more) complete apsidal period(s) (see e.g.
  \citealt{Deeter1987}). But so far, the apsidal period of \oao\ is
  not known.

  For the time covered by the \emph{INTEGRAL} GPS observations we
  find that the centre of eclipse is earlier than our
  $T_{90}$ by $\Delta T=0\fd25\pm 0\fd05$. Using the values in
  Table~1 of \citet{Chak1993} we estimate that also in 1991/92 the
  eclipse was earlier than $T_{90}$ by $\Delta T=0\fd1\pm 0\fd05$.
  In principle, such observed time differences $\Delta T$ can be
  used to find the longitude of periastron. To first order $\Delta T
  = T_\mathrm{ecl} - T_{90} = -(P_\mathrm{orb}/\pi)~e \cos \omega$
  \citep{Deeter1987}, where $\omega$ is the longitude of periapsis
  and $e$ the eccentricity. From the change of $\omega$ the rate of
  apsidal advance (and the apsidal period) can be found.
  Unfortunately, neither the data of \citet{Chak1993} nor our own
  data are accurate enough to yield a meaningful constraint.

   \begin{table}
      \caption[]{Current best set of orbital parameters:
                 $T_{90}$ and period as estimated in this work,
                 $a\sin~i$, eccentricity and $\omega$ taken from \citet{Bild1997}.}
         \label{OrbitalParameters}
         \centering
         \begin{tabular}{l r @{.} l l}
         \hline \hline
                Parameter & \multicolumn{2}{c}{Value} & Unit \\
         \hline
                Epoch ($T_{90}$)  & 52663&893(10)   & MJD \\
                Period            &    10&44812(13) & d \\
                $a\sin~i$         &   106&0(5)      & lt-sec \\
                eccentricity      &     0&104(5)    &   \\
                $\omega$          &    \multicolumn{2}{c}{93(5)~~~~~~~} & deg \\
         \hline
         \end{tabular}
   \end{table}


\begin{figure}
  \includegraphics[width=1.0\columnwidth]{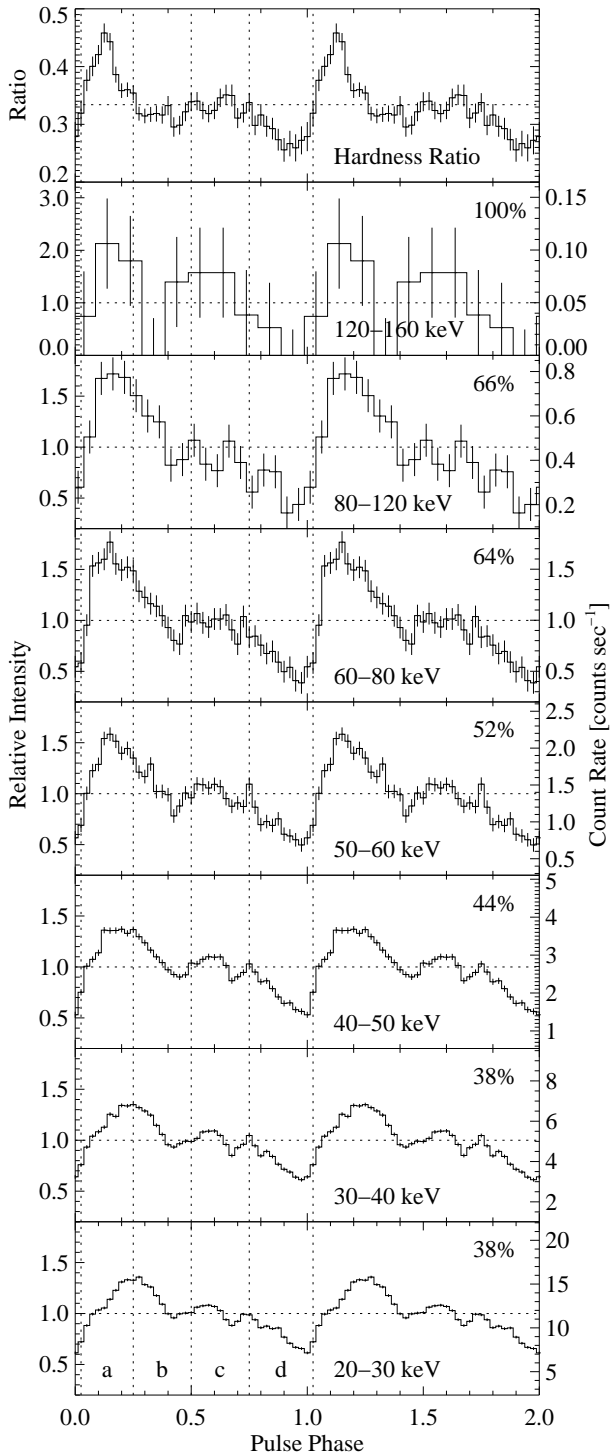}
  \caption{Energy-resolved pulse profiles. The relative intensity
  scale is normalised to the average count rate level (horizontal
  dashed line) in the corresponding energy interval. The percentage
  values indicate the pulsed fraction. The pulsed fraction of the
  highest energy interval was set to 100\%, as the negative minimum
  flux values were set to zero. The top panel shows the hardness
  ratio (see text). The vertical dashed lines denote the four phase
  intervals for which spectra were extracted. The pulse phase limits
  of the four intervals are: $\mathrm{a}=0.025-0.250$,
  $\mathrm{b}=0.25-0.50$, $\mathrm{c}=0.50-0.75$ and
  $\mathrm{d}=0.750-1.025$.}
\label{FigPulseProfiles}
\end{figure}

\begin{figure}
\includegraphics[width=1.0\columnwidth]{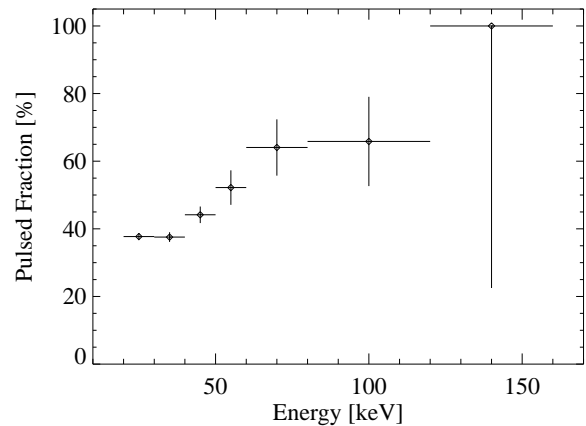}
  \caption{Variation of the pulsed fraction with energy.
  Horizontal bars denote the energy range for which the pulsed fraction
  was calculated.}
\label{FigPulsedFraction}
\end{figure}

\subsection{Energy-resolved pulse profiles}

  With the above determined pulse periods, pulse profiles in seven
  different energy intervals (covering the range $20-160$\,keV) of
  720 SCWs were produced and superposed (using the sharp rise of the
  main pulse as phase reference - we call this pulse phase~0). The
  final energy-resolved pulse profiles are displayed in
  Fig.~\ref{FigPulseProfiles}. The pulsation is clearly detected up
  to the highest energy range ($120-160$~keV).

  The profiles consist of at least three components, the amplitudes
  of which decrease during the course of the pulse. For easier
  comparison the profiles for the different energy intervals are
  shown in the same relative scale (except for the energy range
  $120-160$~keV, where the plot needs a different scale due
  to the higher pulsed fraction and the larger errors). The pulsed
  fraction -- indicated as percentage in Fig.~\ref{FigPulseProfiles}
  -- increases linearly with energy in the range
  $30-70$\,keV (Fig.~\ref{FigPulsedFraction}). We calculated the
  pulsed fraction from the minimum and maximum flux values in the
  corresponding energy interval as
  $f_\mathrm{pulsed}=(F_\mathrm{max}-F_\mathrm{min})/(F_\mathrm{max}+F_\mathrm{min})$.
  This is -- at least in the case of sinusoidal pulse profiles --
  equivalent to the standard definition
  $f_\mathrm{pulsed}=F_\mathrm{pulsed}/(F_\mathrm{pulsed}+F_\mathrm{constant})$,
  calculated from the values of the pulsed and the constant flux
  parts. It should be noted that negative $F_\mathrm{min}$ values
  were set to zero to avoid calculated pulsed fractions larger than
  100\%, which is the case for the highest energy interval shown in
  Fig.~\ref{FigPulseProfiles}.

  Also obvious from the plot is that the first maximum becomes
  asymmetric at higher energies, which indicates a harder spectrum
  in phase interval \emph{a} (defined in
  Fig.~\ref{FigPulseProfiles}). The top panel of
  Fig.~\ref{FigPulseProfiles} shows the hardness ratio calculated as
  ratio of the photon fluxes in the $40-160$\,keV band to those in
  the $20-40$\,keV band.


\section{Spectral modelling}
\subsection{Introduction}
\label{SpectralModeling_Introduction}

\begin{figure}
\includegraphics[width=1.0\columnwidth]{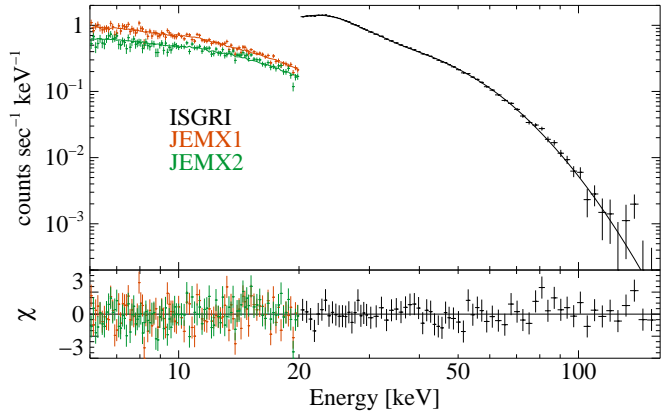}
  \caption{Phase-averaged broadband spectrum for ISGRI, JEM-X1,
  JEM-X2. The fitted model was HIGHECUT.
  To avoid artefacts in the residuals of the fit, the
  transition to the high energy cutoff was smoothed with an
  additional Gaussian absorption. The fit parameters are given in
  Table~\ref{TabFitParams}.}
\label{FigBbSpectrum}
\end{figure}

\begin{table*}
  \caption[]{Table of fit parameters for the phase-averaged broad band
             spectrum shown in Fig.~\ref{FigBbSpectrum} for two
             different spectral models (see text). A systematic
             error of 2\% was assumed. The uncertainties given are
             at 90\% confidence level for single parameter errors
             ($\chi^2_\mathrm{min} + 2.7$).}
  \label{TabFitParams}
  {\centering
  \begin{tabular}{l l l c c c c c c c}
    \hline \hline
           Compon. & Parameter                      & Unit                           & Model A               & Model B               & Ref.~a & Ref.~b      & Ref.~c & Ref.~d & Ref.~e\\
    \hline
           phabs   & $N_\mathrm{H}$ (JEM-X1, ISGRI) & $10^{22}$ cm$^{-2}$            &                       &   $0$ (frozen)        &   1.29 &  14.1        & 41    & 12.7 &  7.2 \\
           phabs   & $N_\mathrm{H}$ (JEM-X2)        & $10^{22}$ cm$^{-2}$            &                       &   $18.4^{+3.7}_{-3.6}$&        &              &       &      &      \\
           powerlaw  & $\Gamma$ (JEM-X2, ISGRI)     &                                &   $1$ (frozen)        &   $1.27{\pm 0.05}$    &   1.07 &              & 1.0   & 0.83 &  0.60 \\
           powerlaw  & $\Gamma$ (JEM-X1)            &                                &   $1.3$ (frozen)      &   $1.27{\pm 0.05}$    &        &              &       &      &   \\
           powerlaw  & normalisation (1~keV)        & (keV cm$^{2}$ s)$^{-1}$   &  $0.055^{+0.003}_{-0.002}$ &   $0.12{\pm 0.02}$    &   0.06 &              & 0.046 &      &   \\
           highecut  & Cutoff energy                & keV                            &  $21.6^{+1.0}_{-1.4}$ &  $24.4{\pm 0.9}$      &  12.82 &              &       & 13.0 &  $<$5.0 \\
           highecut  & Folding energy               & keV                            &  $19.0{\pm 0.2}$      &  $21.2{\pm 0.5}$      &  29.37 &              &       & 21.2 &   17.0 \\
           gaussian  & Fe line centre               & keV                            &                       &                       &   6.65 &  6.4 / 7.1   & 6.4   & 6.47 &   6.60 \\
           gaussian  & Fe line width $(\sigma)$     & keV                            &                       &                       &   0.40 &  0.2 / 0.3   & 0.044 & 0.42 &   0.24 \\
           gaussian  & Total flux in Fe line        & $10^{-3}$ (cm$^{2}$ s)$^{-1}$  &                       &                       &   2.77 &  0.74 / 0.25 & 0.8   &      &   3.0 \\
    \hline
           energy range  &                          & keV                            & 6--160                & 6--160                &  3--100 & 0.5--10      & 4--8 & 1--100 & 3--40 \\
           red.~$\chi^2$ &                          &                                & 1.00                  & 0.94                         \\
           DOF           &                          &                                & 258                   & 253                          \\
    \hline
  \end{tabular}
  }
    ~\\
    $^{\mathrm{a}}$ \citet{Baykal2000};
    $^{\mathrm{b}}$ \citet{Audley2006}, non eclipse spectrum, 2 Fe lines;
    $^{\mathrm{c}}$ \citet{2002ApJ...573..789C};
    $^{\mathrm{d}}$ \citet{Orland1999};
    $^{\mathrm{e}}$ \citet{1990PASJ...42..785K}

\end{table*}

  Due to calibration uncertainties below 20\,keV and low flux above
  160\,keV the useable range of the ISGRI spectra was $20-160$~keV.
  The JEM-X spectra were used in the $6-20$~keV range, which we
  consider as the most reliable range. All spectra are background
  subtracted.

  We used the following components for our spectral models:
  An absorbed power law model ($\exp[-\sigma (E) N_\mathrm{H}]
  E^{-\Gamma}$, XSPEC model \texttt{phabs*powerlaw} -- in the
  following ABSPL), which influences only the JEM-X part of
  the spectra. The second component was an exponential
  cutoff power law model ($E^{-\Gamma}\exp[-E/E_\mathrm{f}]$, XSPEC
  model \texttt{cutoffpl} -- in the following CUTOFFPL) or a power
  law with high energy cutoff
  ($E^{-\Gamma}\exp[-(E-E_\mathrm{c})/E_\mathrm{f}]$ if
  $E>E_\mathrm{c}$, else $E^{-\Gamma}$; XSPEC model
  \texttt{powerlaw*highecut} -- in the following short HIGHECUT).
  These components are mainly determined by the ISGRI part
  of the spectra.

  Being not able to fit the spectra with CUTOFFPL plus
  cyclotron absorption as described by \citet{Orland1999}, we used
  in our analysis the HIGHECUT model (Model A in
  Table~\ref{TabFitParams}). In order to avoid artefacts in the
  residuals usually produced by the XSPEC \texttt{highecut} model at
  $E_\mathrm{c}$ \citep{Kretsch1997}, the function was smoothed with
  a Gaussian absorption \texttt{gabs} as successfully exercised by
  \citet{Coburn2002}. The centre of this Gaussian was set to
  $E_\mathrm{c}$ plus 1\,keV (from experience with other spectra and
  confirmed by the fitting procedure). The width $\sigma$ and the
  depth $\tau$ of the Gaussian were found to be 2.2\,keV and 0.9,
  respectively, and were subsequently fixed at these values.

  The photon index $\Gamma$ for the ISGRI part of the phase-averaged
  spectra (Fig.~\ref{FigBbSpectrum}) was fixed to unity, which is
  justified by the JEM-X2 part of the spectrum and is consistent
  with other spectral fits reported in literature (see
  Table~\ref{TabFitParams}). The JEM-X1 spectra show a
  significantly different photon index of 1.3 (the
  errors for all $\Gamma$ values are 0.05). There are several
  possible explanations for this behaviour: First, the JEM-X1
  spectra cover the orbital phases from 0.25 to 0.65, while the
  JEM-X2 spectra are mainly located in the phase intervals
  $0.1-0.25$ and $0.65-0.8$. Second, they originate from different
  time intervals: JEM-X2 from MJD $52\,671-52\,869$ and JEM-X1 from
  MJD $53\,232-53\,795$. A third possibility might be the existence
  of uncertainties in the spectral calibration of the two
  instruments.

  This difference in the JEM-X spectra could also be well
  fitted by an additional photon absorption in the JEM-X2 spectra
  (ABSPL, see Sect.~\ref{SpectralModeling_Introduction}). Model B
  in Table~\ref{TabFitParams} shows the parameters in this model,
  which is formally Model A multiplied by the XSPEC model
  \texttt{phabs}. The photon index $\Gamma$ was kept identical for
  all 3 spectral parts (ISGRI, JEM-X1, JEM-X2). No absorption was
  assumed for the ISGRI and JEM-X1 part, while for JEM-X2 cold
  matter absorption was fitted. The fit led to a $N_\mathrm{H}$ of
  $(18.4\pm3.6)\times 10^{22} \mathrm{cm}^{-2}$, which is within the
  range of other literature values shown in
  Table~\ref{TabFitParams}. This finding could imply that the
  spectra near the eclipse are more absorbed than in the centre of
  the orbital profile. On the other hand it could also mean that at
  least part of the strong variability of \oao\ is due to variable
  absorption, which was higher during the JEM-X2 part of the
  observation.


\subsection{Pulse-phase resolved spectroscopy}

\begin{figure*}
  \includegraphics[width=2.0\columnwidth]{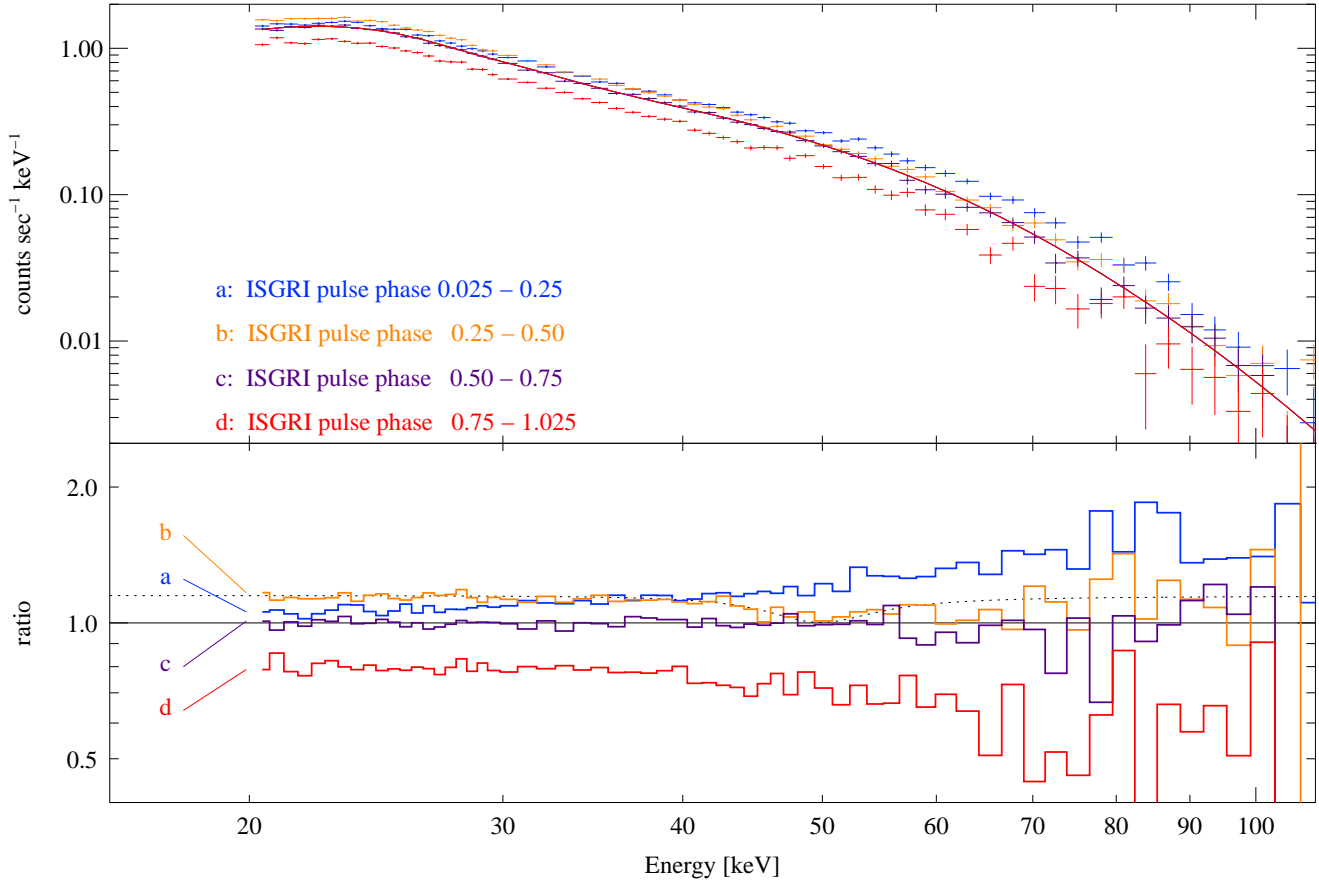}
  \caption{Phase-resolved ISGRI spectra for the four phase intervals
  as defined in Fig.~\ref{FigPulseProfiles}. The solid line in the
  upper panel is the model fit of the phase-averaged spectrum as shown
  in Fig.~\ref{FigBbSpectrum} (model A of Table~\ref{TabFitParams}).
  To enhance the differences between the spectra of the four phase
  intervals the lower panel gives the ratios between the four spectra and
  the model of the phase-averaged spectrum in a logarithmic scale.
  The dotted line in the lower panel corresponds to a fit of
  spectrum {\it b} with a line feature at 49.3\,keV.}
\label{Fig4PhSpectrum}
\end{figure*}

  We have investigated ISGRI spectra from the four pulse phase
  intervals indicated in Fig.~\ref{FigPulseProfiles}. The spectra
  are shown in Fig.~\ref{Fig4PhSpectrum}. We fitted all data
  using the HIGHECUT model, keeping the power law index
  fixed to $\Gamma=1.0$. The results are given in
  Table~\ref{TabPhaseFitParams}.

\begin{table*}
  \caption[]{Fit parameters for the spectral model HIGHECUT,
             Model A in Table~\ref{TabFitParams}.

             In the last line the fit parameters for the second
             phase interval are given for the same model, but
             multiplied by a cyclotron absorption$^{\mathrm{a}}$.
             The uncertainties given are at 90\% confidence level
             for single parameter errors ($\chi^2_\mathrm{min} +
             2.7$). No systematic errors were considered.}

  \label{TabPhaseFitParams}
  \centering
  \begin{tabular}{r @{ -- } l c c c c c c c}
    \hline \hline
           \multicolumn{2}{c}{Phase} &  Red.~$\chi^2$ & DOF & $\Gamma$ & Normalisation at 1~keV              &  Cutoff Energy & Folding Energy \\
           \multicolumn{2}{c}{}      &                &     &          & [$10^{-2}$ (keV cm$^{2}$ s)$^{-1}$] &  [keV]         & [keV]          \\
    \hline
           0.025 & 0.25   &  0.99 & 68 &  1  &  6.1$_{-0.3}^{+0.5}$ & 20.1$_{-0.3}^{+0.6}$ & 21.4$_{-0.3}^{+0.6}$  \\
           0.25  & 0.5    &  1.30 & 68 &  1  &  6.2$_{-0.2}^{+0.4}$ & 22.0$_{-0.3}^{+0.5}$ & 18.1$_{-0.3}^{+0.5}$  \\
           0.5   & 0.75   &  1.02 & 68 &  1  &  5.4$_{-0.2}^{+0.5}$ & 22.1$_{-0.3}^{+0.6}$ & 18.7$_{-0.3}^{+0.6}$  \\
           0.75  & 1.025  &  1.26 & 68 &  1  &  4.2$_{-0.1}^{+0.2}$ & 23.2$_{-0.3}^{+0.6}$ & 17.1$_{-0.3}^{+0.6}$  \\
    \hline
           0.25  & 0.5    &  1.03 & 68 &  1  &  6.4$_{-0.2}^{+0.5}$ & 20.9$_{-0.9}^{+1.9}$ & 19.1$_{-0.3}^{+0.5}$  \\

    \hline
  \end{tabular}

    ~\\
    $^{\mathrm{a}}$ The cyclotron line parameters
             (XSPEC model \texttt{cyclabs}) were: energy 49.3\,keV,
             width 5.4\,keV, depth 0.136. These parameters were
             fixed after fitting them to a spectral range of
             $20-60$\,keV, so that for the fit parameters shown in
             the last line also a DOF value of 68 is valid.
\end{table*}


\subsection{Search for cyclotron resonance scattering features}

  As cyclotron resonance scattering features are observed in several
  pulsars (see discussion in Sect.~\ref{summary_timing_analysis}) we
  tried to find possible evidence of a cyclotron line feature. We
  therefore calculated the ratio of the spectra from the four phase
  intervals to the phase-averaged spectrum (model A of
  Table~\ref{TabFitParams} with ISGRI parameters). In this way,
  differences between the individual spectra and features within
  these spectra are enhanced. The four individual spectra and the
  ratios are shown in Fig.~\ref{Fig4PhSpectrum}. While spectrum c
  (pulse phase $0.5-0.75$) is nearly identical to the pulse-phase
  averaged spectrum (the ratio in the lower panel of
  Fig.~\ref{Fig4PhSpectrum} is around 1.0), spectrum a (phase
  $0.025-0.25$) is clearly harder and spectrum d (phase
  $0.75-1.025$) is clearly softer. Spectrum b (phase $0.25-0.5$)
  shows a slight ``peculiarity'': there seem to be two components to
  the spectrum with a transition between 45\,keV and 55\,keV, or
  alternatively, a depression around these energies. Formally, a
  cyclotron line can be fitted with a centroid energy of 49.3\,keV.
  The $\chi^2$ is reduced to 0.9 (DOF\,=\,41) from 1.2
  (DOF\,=\,44), resulting in an F-Test value of 1\,\%,
  which is usually not considered as sufficient to state
  the detection of a cyclotron absorption line -- also one
  should keep in mind that it is in general not correct to use an
  F-Test to test for the presence of a line \citep{Protassov2002}.


\section{Summary and Discussion}
\label{Summary_and_discussion}

\subsection{The orbital profile}

  In the averaged light curve (Fig.~\ref{FigOrbitalLightcurve}) we
  found a dip at phase interval 0.55, which is confirmed by ASM
  measurements (Fig.~\ref{FigOrbitalLightcurveASM}) taken in
  the same time interval as the INTEGRAL observations, but
  homogeneously sampled. The dip is more pronounced in our
  \emph{INTEGRAL} data, but we cannot exclude that this is
  an effect from the inhomogeneous sampling in the \emph{INTEGRAL}
  data (see dashed line in both figures) with a minimum of
  measurements at the minimum of the profile. In order to evaluate
  the effects of the sampling we ran a statistical simulation, used
  to produce the ``error bars'' in Fig.~\ref{FigOrbitalLightcurve}.
  These bars show the range between the minimum and maximum values
  evaluated from 10\,000 profiles calculated from 30 randomly chosen
  SCWs per phase bin. The dip around orbital phase 0.5 is always
  present, in line with the ASM results.

  We note that the average profile should not be
  seen as a template for individual light curves, it rather
  seems to be the result of a random process. Some individual light
  curves have a maximum in the phase interval $0.2-0.5$, others (and
  more) have a maximum in the interval $0.6-0.85$. So the strong
  variability of the source may be weakly correlated with the
  orbital phase in such a way that the chance for a ``low state'' is
  quite high at phase~0.55.

  The physical origin of this orbital modulation is still unclear.
  Topics to be discussed in this context should comprise
  absorption, scattering, a (partial) blockage of the line of sight
  to the neutron star, or variations of the accreted wind. All of
  these mechanisms will be subject to further investigations.

  As the pulsar can be considered as a point source compared
  to its B supergiant companion, the slope of the ingress and egress
  indicates a soft transition from the stellar atmosphere to the
  stellar wind region with a decreasing optical thickness for the
  radiation emitted by the pulsar. The density profile of the outer
  regions of the supergiant was modelled by \citet{Bulik2005} with an
  exponential density distribution. They fitted the modelled
  absorption to the observed folded light curve, which was composed
  from 480 \emph{INTEGRAL} SCWs. This model did not account for the
  asymmetric shape of the orbital profile and should be
  refined in a further analysis.

\subsection{Timing analysis}
\label{summary_timing_analysis}

  The \emph{INTEGRAL} observations extend the long-term monitoring
  of \oao\ and cover a period with a torque reversal from strong
  spin-down to strong spin-up. We have compiled all available pulse
  period measurements from the literature and, where necessary,
  converted the values to the intrinsic neutron star spin period. We
  find that the long-term spin-up is continued and determine a mean
  rate of $\dot{P}=-1.24\times 10^{-9} s\,s^{-1}$. The
  spin-down/spin-up behaviour observed by \emph{INTEGRAL}, together
  with the historical data imply a quasi-periodic variation
  of the pulse period with a period of roughly 5 years, superimposed
  on the long-term mean spin-up.

  The orbital elements were determined from pulse timing
  analysis. Combining our results with previous data we determine an
  orbital period with improved accuracy:
  $P_\mathrm{orb}=10\fd448\,12\pm 0\fd000\,13$. We find a difference
  between $T_{90}$ and the centre of eclipse of $(0\fd3\pm 0\fd1)$,
  which is consistent with the previously determined value, and does
  not allow a conclusion about a possible apsidal advance.

  The long-term evolution of the spin period is similar to
  that of \object{Her~X-1}. In this object a quasi-periodic
  spin-up/down variation is found with a period of about 5\,yr, (as
  in \oao) superimposed on a long-term spin-up \citep{Staubert2006}.
  Since in Her~X-1 the period variations are correlated
  with the 35\,d turn-on behaviour and the quasi-periodically
  repeating anomalous low states, this phenomenon is generally
  connected to the precessing accretion disk and an associated
  variation in the mass accretion onto the neutron star.

  In contrast to Her~X-1, however, \oao\ is a high mass, wind
  accreting system and we have no direct evidence of an accretion
  disk. Also no correlation between flux and period change is
  observed \citep{Inam2000}. So the conditions are probably
  different compared to the low mass X-ray binary Her~X-1. \oao\ is
  probably better compared to other high mass X-ray binaries (HMXB),
  e.g. \object{Cen~X-3} and \object{Vela~X-1}. The long-term
  frequency histories presented by \citet{Bild1997} for several
  pulsars, show a secular spin up for Cen~X-3 with a weak quasi
  periodic modulation of about 5 to 8 years. For Vela~X-1, on the
  other hand, a secular spin down is shown, but with a torque
  reversal around MJD~44\,000 (spin up to spin down) and possibly also
  around MJD~49\,000 (spin down to spin up). Between these dates also
  a quasi periodic characteristics of the period evolution
  with a period around 5 years might be inferred. We note here that
  these quasi-periodic pulse period variations, seen in several
  accreting X-ray pulsars, have not found much attention in the
  literature.

  In addition to the long-term quasi-periodicity we have found
  variations of the pulsar period on time scales shorter than the
  orbital period (residuals in Fig.~\ref{FigPulseDelay}).
  \citet{Bild1997} have discussed power spectra of torque
  fluctuations for several objects, including \oao . While Her~X-1
  and Vela~X-1 showed a white noise power density spectrum, Cen~X-3
  shows red noise. For \oao\ \citet{Bild1997} also noted a red noise
  spectrum. But a closer look at the power density spectrum shows a
  slight increase in power density for frequencies above the orbital
  frequency of about $1 \times 10^{-6}$~Hz.  \oao\ has the highest
  power density at high frequencies of all pulsars presented in
  \citet{Bild1997}, which is consistent with our finding of
  pulse period changes on time scales much shorter the orbital
  period. Also \citet{Blondin1990} expect fluctuations in the
  spin-up/spin-down on short time scales as a result of changes in
  the sign and magnitude of the accretion torque.

\subsection{The pulse profile}

  The energy-resolved pulse profiles show a complex
  structure with one main peak followed by at least two somewhat
  weaker peaks. The spectrum of the rising part of the main peak is
  clearly harder than in the other parts of the pulse.

  The change of spectral hardness with pulse phase is
  commonly observed. Below 10\,keV this behaviour might be modelled
  by a variation of an absorbing column density $N_\mathrm{H}$
  \citep[e.g. Cen~X-3,][]{Santa1999a} but for energies above
  $10-20$\,keV cold absorption plays no role anymore. A mixture of
  different emission regions, as proposed by \citet{Kraus1996} to
  explain energy-resolved pulse profiles of Cen~X-3, might explain
  the observed profiles, but such detailed theoretical modelling is
  beyond the scope of this paper.

  As in many other accreting pulsars, e.g. \object{4U~0115+63}
  \citep{Tsyg2007} or \object{GX 1+4} \citep{Ferrigno2007}, the
  pulsed fraction increases with energy. \citet{Tsyg2007} point out
  that in the accretion column the region emitting harder photons is
  closer to the neutron star surface and therefore can be more
  obscured during certain pulse phases than the region higher up
  emitting photons of lower energy.

\subsection{Spectral analysis}

  The broad band spectrum, composed from ISGRI and JEM-X spectra
  could be modelled by the HIGHECUT model. We could not find an
  indication for a cyclotron absorption in the phase-averaged
  spectrum.

  \citet{Blondin1990} have done hydrodynamic simulations of the
  accretion by a neutron star from stellar wind in massive X-ray
  binaries. Their work points to episodic changes of the mass
  accretion rate, which results in corresponding changes of the
  X-ray luminosity on a time scale of hours. They also find a
  variation of integrated column density, which varies with orbital
  phase in such a way that it is highest near eclipse and lowest at
  orbital phase 0.5. This is in good agreement with our findings for
  the JEM-X1 and JEM-X2 spectra, which show a higher absorption near
  eclipse.

  Phase-resolved spectra were produced from ISGRI observations for 4
  phase intervals. These spectra show significant differences. Phase
  $0.025-0.25$, covering the rising part of the main pulse peak,
  shows the hardest spectrum, while phase $0.75-1.025$ has the
  softest spectrum. Phase $0.5-0.75$ fits nearly exactly the
  phase-averaged spectrum, while phase $0.25-0.5$ shows a feature at
  about 50\,keV, which is reminiscent of a cyclotron absorption
  line, but with a significance too weak to prove its existence.

  Apart from not being able to detect a cyclotron line in
  our data, it would not be unlikely to find such a line in the
  spectrum of phase interval $0.25-0.5$, as it is common to several
  pulsars that cyclotron lines are visible only -- or at least most
  prominent -- in the falling part of the main pulse, e.g.
  4U~0115+63 \citep{Santa1999}, \object{GX~301-2}
  \citep{Kreyken2004}, Cen~X-3 \citep{Burderi2001}, Vela~X-1
  \citep{Kreyken2002} and possibly also GX~1+4, which shows a
  likewise weak absorption feature at 34\,keV \citep{Ferrigno2007}.

\section{Conclusions}

  We have presented a spectral and timing analysis of
  \emph{INTEGRAL} observations of \oao , which span a time range of
  about 3.5 years. Though we found our results to be compatible with
  similar observations of other high mass X-ray binaries, we state
  that a consistent model for \oao\ is still missing. Such a model
  should explain the nature of the strong variability and its
  possibly weak correlation with the orbital phase. Also the origin
  of the fluctuations of the pulse period with its observed long-term
  quasi periodic behaviour should be addressed by this model.
  Detailed analysis is also required to explain the rather complex
  structure of the pulse profiles. Further observations with
  continuous coverage of \oao\ would be required to settle the
  question of a potential spectral feature in the falling part of
  the main peak of the pulse profile.

\begin{acknowledgements}
      We thank Mark Finger for complementing the publicly
      available data on pulse frequency. Part of this work is
      supported by the German Space Agency (DLR) under contracts
      50\,OG\,9601 and 50\,OG\,0501.
      We also thank the anonymous referee for valuable comments and
      suggestions.
\end{acknowledgements}

\bibliographystyle{aa}
\bibliography{jb_bib}

\listofobjects
\end{document}